\documentclass[10pt]{iopart}

\usepackage{graphicx}
\usepackage{mathptmx}
\usepackage{etoolbox}
\usepackage{iopams}
\usepackage{cite}
\usepackage{makecell}
\usepackage{epstopdf}
\begin{document}

\title[]{Controlling the electric force on a dust particle during the afterglow of a plasma at a higher gas pressure}

\author{Neeraj Chaubey$^1$ and J. Goree$^2$}

\address{$^1$Physics and Astronomy Department, University of California Los Angeles, Los Angeles, California 90095, USA}
\address{$^2$Department of Physics and Astronomy, University of Iowa, Iowa City, Iowa 52242, USA}

\ead{nchaubey@physics.ucla.edu}
\vspace{10pt}

\begin{abstract}
When dust particles are immersed in a plasma, and the power that sustains a plasma is terminated, the charge of dust particles will change in the early afterglow, as electrons and ions gradually diminish in number. The possibility of controlling this charge, along with the electric force acting on the particles in the late afterglow, has earlier been demonstrated at a low gas pressure of 8 mTorr. Here, it is confirmed experimentally that controlling particles is possible also at a higher gas pressure of 90 mTorr, in a capacitively coupled radio-frequency plasma (CCP). A timed application of a DC electric field during the afterglow is a key element of this control scheme. Analyzing the experimental results, the electric force in the late afterglow was determined by comparing measurements of particle velocity to a prediction made by integrating the equation of motion, taking into account gas friction. In addition to applying friction to dust particles, gas also slows the drifting motion of electrons and ions, reducing their energy during the afterglow, but nevertheless we find that dust particles become charged in the afterglow so that one can apply an electric force to them that is comparable to the gravitational force, even at a higher pressure than had previously been demonstrated. This result extends the parameter range for which it is expected that particle contamination in semiconductor manufacturing can be mitigated by controlling charge and forces during the afterglow. Because of the way that forces scale with particle size, it is expected that submicron particles can be controlled even more easily than the larger spheres in the present experiment.
\end{abstract}

\vspace{2pc}
\ioptwocol

\section{Introduction}
A dusty plasma is an ionized gas containing small particles of solid matter, which become charged by collecting electrons and ions \cite{Shukla_2002, Fortov_2004, Tsytovich_2008, Goreebook_2008, Goertz_1989, Mangilal_2021b, Thomas_2016, Mangilal_2020, Samsonov_2000, Nunomura_2000, hebner_2002, Nosenko_2003,Hartmann_2005, Feng_2007, Knapek_2007, Ruhunusiri_2011,Dharodi_2014,tiwari2014kelvin,  Kostadinova_2018, kaur_2015, Dharodi_2020b, knapek2022compact, Mengel_2023, ramkorun2024comparing, li2025progress, dharodi2023ring, mengel2025equivalent, gogoi2024observation, shakoori2024phase, ramkorun2024introducing, mccabe2025experiments, kumar2025elastic}. Until recently, experiments with dusty plasmas were mostly done with steadily powered laboratory plasmas. In contrast, there has been an increasing interest in dust particles in afterglow plasmas \cite{sikimic2014dynamics, stefanovic2017influence, chen2019, platier2020, chen2022, staps2021, collins1996particle, Barkan_1995,setyawan2003characterization, merlino_2016, meyer_2016, merlino_2018,Ivlev_2003, couedel_2006, couedel_2008, couedel_2009, Schweigert_2012, saxena_2012, worner_2013, denysenko_2013, denysenko_2016, denyesenko_2018, Denysenko_2021, Van_2022, Minderhout_2019, Minderhout_2020, Sharma_2020,dhawan2020enhancing, Minderhout_2021, denysenko2022modeling, staps2022review, van_2023, chaudhuri2023particle,abuyazid2023charge, beckers2023role, choudhary2023magnetized, knapek2024void, chaudhuri2025transient}. 

An afterglow is a condition after the power that sustained the plasma is turned off. During the afterglow, electrons and ions gradually depart the chamber, over the course of milliseconds. During this time, a dust particle’s charge can change significantly \cite{Neeraj_1, Neeraj_2, Neeraj_3, Neeraj_4,Neeraj_5, Neeraj_6, Ivlev_2003, filatova2011, Couedel_2022, zhang2024jump,van2024influence, denysenko2024plasma}. 

For a capacitively coupled radio-frequency plasma (CCP), experimental results from our laboratory have shown that in the afterglow, a dust particle can develop a charge that can be quite large \cite{Neeraj_1, Neeraj_2, Neeraj_3, Neeraj_4,Neeraj_5, Neeraj_6}. Therefore, the particle can experience a large electric force when there is an ambient electric field. This force was the product of the dust particle’s residual charge $Q_\mathrm{res}$ and the ambient electric field, during the late afterglow\cite{Neeraj_1}.  The magnitude of this electric force could be made as large as the gravitational force $M_{d}\,g$, depending on the voltage on a negatively biased lower electrode during the afterglow. Here, $M_d$ is the dust particle’s mass and $g$ is the acceleration of gravity. Those results were for a rather low Argon pressure of 8 mTorr (1.066\,Pa) \cite{Neeraj_1}.

In this paper we report experimental results showing that a comparable large electric force can be obtained with a higher gas pressure of 90\,mTorr\,(12 Pa). Our experiment was performed in the same chamber, with the same kind of dust particles as in \cite{Neeraj_1}. Instead of allowing the lower electrode to remain at the negative DC bias as in \cite{Neeraj_1}, we applied a positive DC potential to the lower electrode at a specific time in the early afterglow, as in \cite{Neeraj_4}, so that dust particles near that electrode collect electrons, instead of ions. Differently from those earlier experiments, here we used a higher gas pressure, which has two consequences: slowing the drifting motion of electrons and ions during the early afterglow, and applying drag to the dust particles.
 
The higher gas pressure in the present experiment may make these results of interest for semiconductor
manufacturing. In semiconductor manufacturing, contamination due to defects caused by particles is a significant problem, with particles falling onto the wafer during the afterglow. For that reason, there is interest in using electric forces, as in the current paper, to manipulate the motion of dust particles
in the afterglow \cite{Neeraj_5}. The gas pressure typically can exceed 8\,mTorr by an order of magnitude in lithography machines producing vacuum ultraviolet light \cite{chaudhuri2025transient}. Even higher gas pressures are common for plasma etching and especially chemical vapor deposition (plasma enhanced CVD). For these reasons, we are motivated to extend the demonstrated range for our afterglow method of particle control, to pressures higher than the 8\,mTorr that we previously reported \cite{Neeraj_1, Neeraj_2, Neeraj_3, Neeraj_4, chaubey2023dust, Neeraj_5, Neeraj_6}.

To measure the electric force $F_E$ acting on dust particles in the afterglow, our method is to observe them falling using a high-speed video camera, and analyze the images to obtain a time series of velocities. At the higher gas pressure of 90 mTorr in the present experiment, our analysis of the velocity must include the effect frictional drag on the dust particle due to gas, as we will do in this paper.  This approach requires a numerical solution of the dust particle’s equation of motion, which has one free parameter, corresponding to the electric force $F_E$ acting on the dust particle. By adjusting that free parameter $F_E$ to obtain agreement with experimental results for the velocity time series, we obtain its value in the afterglow.

We find that at a pressure of 90 mTorr, we were able to apply an electric force that was comparable to that of gravity. We found that $F_E /M_d\,g$ was in the range 0.42 to 0.72,  when applying a modest electric potential of +150\,V to the lower electrode, for microspheres of diameter 2$\,R_d\,=\,8.69\,\mu$m. 

Scaling of forces depends on particle size in such a way that we expect that for smaller particles the electric force should be even stronger in comparison to that of gravity.  The ratio of $F_E / M_d\,g$ should scale as $R_d^{-2}$, since the charge scales as the first power of $R_d$. This scaling suggests that smaller dust particles, like those that are typically of interest in the semiconductor industry, can be manipulated by the timed application of electric fields of even a modest magnitude.

\section{Experiment method}
\subsection{Apparatus}
For the present experiment at a pressure of 90 mTorr, we used the same apparatus as in \cite{Neeraj_4}, where we operated the plasma at 8\,mTorr. This apparatus, sketched Fig.\,1, allows us to observe motion of particles during the afterglow to determine the forces acting them. It also allows us to apply a DC bias to the lower electrode at a specified delay time, during the afterglow. We will briefly review that apparatus.
 
The vacuum chamber had an inside diameter of 20.15\,cm. The lower electrode of 16.25 cm diameter was connected to a radio-frequency power supply through a coupling capacitor $C_\mathrm{coupl}$\,=\,61.2\,nF. All metal surfaces in the chamber were grounded, except for the lower electrode. 

An argon plasma was started by a steady 13.56\,MHz radio-frequency (RF) power. We operated with a peak-to-peak voltage of 91\,V and DC self-bias of -30\,V, measured using a Tektronix P5100A 100X probe. Under these conditions, the dust layer was centered above the lower electrode. To produce an afterglow, the RF power was switched off at $t$\,=\,0, using a signal from a gate generator that also triggered other electronics. These included a delay generator, which was used to select the time at which a transistor switch was turned on to apply a desired external dc power supply during afterglow.

The dust particles were melamine formaldehyde (MF) microspheres. Their diameter was 8.69\,$\mu$m, i.e., $R_{d\,\mathrm{mfg}}$\,=\,4.35\,$\mu$m, and their density was $\rho_{d\,\mathrm{mfg}}$\,=\,1.51\,g/cm$^3$, as specified by the manufacturer \cite{manufacturer}. We operated the experiment with this large particle size so that we could easily observe them with our camera. For the semiconductor industry, submicron particles would generally be of interest, and we discuss in Sec.\,5.2 how our results can be extrapolated to those smaller sizes.

After introducing the MF particles from above, by agitating a dispenser, they settled into a layer that was electrically levitated above the lower electrode. The particles were negatively charged during steady plasma operation, and lifted upward by the time-averaged electric field of the sheath above the lower electrode. That upward force was balanced, during steady plasma operation, by the downward gravitational force, $M_d\,g$. The value of $M_d\,g$ will be necessary for our calculations, and for that purpose we will use the published \cite{Pavlis_2012} local value of the gravitational acceleration, $g$\,=\,9.804\,m/s$^2$.
 
The particles were illuminated from the side, using a vertical sheet of laser light. The beam from a diode laser, at 671\,nm, was shaped into a thin vertical sheet using a combination of cylindrical and spherical lenses, and it was pointed at the dust layer using mirrors. The particles were imaged by a 12-bit Phantom v5.2 camera, operated at 1000\,frames/sec. The lens, which was fitted with an interference filter to block wavelengths other than that of the laser, provided a spatial resolution so that 0.037\,mm in the dust cloud corresponded to one pixel on the camera’s sensor. The camera was triggered so that it began recording at $t\,=\,0$, when the RF power was turned off. At that time, the particles began falling, in the afterglow. As they fell, the particles were not measurably disturbed by the overall gas flow in the chamber, since we operated at a low flow rate of 0.4\,sccm.

\subsection{Procedure during the afterglow}

The afterglow began at $t\,=\,0$, when the radio-frequency power was turned off. The particles, which had until then been levitated steadily in a single horizontal layer, began to fall. All of them were observed to fall at nearly the same rate.  As they fell, they were imaged by the camera so that their positions could be obtained by analyzing each video image using the moment method \cite{Feng_2007}. We averaged the height of all the particles that were visible in an image, for each frame to reduce random errors. We then used the central-difference method to obtain the time series of the vertical velocity $v_{dz}$ of the dust cloud, calculated at the times of the camera frames.

In the experimental run, we switched off the RF power at $t\,=\,0$, which simultaneously triggered the start of the video recording and activated the delay timer. After the set delay, a DC bias was applied via a transistor switch connected to an external DC power supply.

The DC bias application would be done in two steps, in semiconductor manufacturing, according to our patent-pending process \cite{Neeraj_5,Neeraj_7,Neeraj_8}. The first step would involve applying one DC bias to the lower electrode in the early afterglow while plasma still remained, to control the charge. The second step would be to apply a different bias, later in the afterglow when the plasma is gone but particles remain in the vacuum. This second step would be intended to control the lifting of particles. In the present experiment, we carry out only a single step, for the purpose of demonstrating that even with a higher gas pressure, as compared to our previous experiments, it is possible to manipulate particles in the afterglow. 

In the single step used here, the delay time of $t$\,=\,250\,$\mu$s was chosen to be early enough that electrons would remain in significant quantities. At that delay time, we operated the transistor switch to apply the specified DC bias $V_\mathrm{bias}$ = +150\,V  to the lower electrode. This change in electrode bias changed the nature of the afterglow plasma, especially near the lower electrode. 
 
The electrode became anodic rather than cathodic for $t$\,$>$\,250\,$\mu$s, due to the application of the DC bias. That distinction is significant because it means that electrons in the afterglow were attracted toward the positively biased lower electrode, while ions were repelled. Thus, the dust particles, which in this experiment were in the vicinity of the lower electrode, tended to collect electrons and charge negatively, not positively as would they occur in the afterglow with a cathodic lower electrode. We reported a more detailed discussion of these charging processes in \cite{Neeraj_4}. Even if the particle briefly became positively charged in the first 250\,$\mu$s of the afterglow, it became negatively charged afterward. 
 
In the second step of our patent-pending process to mitigate particle contamination \cite{Neeraj_5, Neeraj_7, Neeraj_8}, the first step of biasing early in the afterglow would be followed by a second step. In that second step, after an additional delay of roughly several milliseconds so that electrons and ions have substantially departed the chamber, the electric field would be reversed by applying a different DC bias to an electrode. With that reversal of the electric field, the dust particles would thereafter experience an upward lifting force, which would slow or prevent the fall of particles to the lower electrode. In the present experiment, we performed only the first step for simplicity, as it is sufficient for the purpose of demonstrating the reversal of charge and the possibility of applying a significant electric force. Thus, in the present experiment, the electric force was downward, and the particles experienced a downward acceleration greater than that of  gravity alone.

\begin{figure}
	\centering

\includegraphics[width=1\columnwidth]{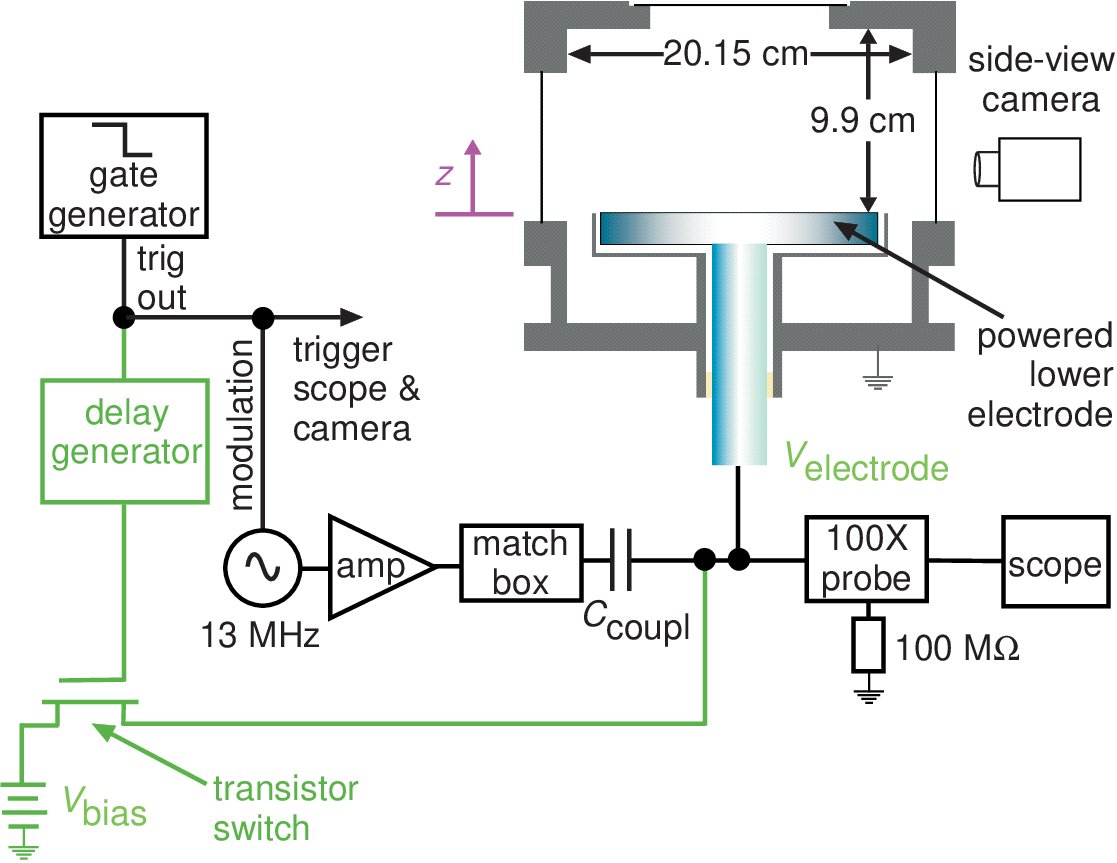}%
\caption{\label{Fig1} Apparatus. Plasma was powered steadily by applying RF power to the lower electrode until $t\,=\,0$. Beginning at $t\,=\,0$, the afterglow began by applying a gate to turn off the RF power. After a delay time, a transistor switch closed so that an external DC power supply applied a potential $V_\mathrm{bias}$ to the lower electrode, overcoming the lingering DC potential on the lower electrode that had until then been sustained by the coupling capacitor $C_\mathrm{coupl}$.  Except for the lower electrode, all the metal surfaces of the chamber surfaces were grounded. A cloud of microspheres, not shown here, was electrically levitated as a single layer 7 mm above the lower electrode for $t\,<\,0$, and as they fell starting at $t\,=\,0$, their motion was recorded by a side-view camera operated at 1000\, frames\,/\,sec.}
\end{figure}

\section{Analysis method}

We will produce theoretical curves for the time series of particle velocity, for comparison to the experiment. This will be done by integrating the equation of motion, taking into account the  forces acting on the falling microspheres. Here we present the equation of motion and describe our method of comparing the theoretical curves to the experimental data, to yield the value of a free parameter, which will be the electric force $F_E$.

\subsection{Equation of motion}
For a small solid particle, as it falls in a plasma afterglow, the equation of motion is 

\begin{equation}
\vec{\dot{v}}_{dz} = -\vec{g} + ({\vec{F}_E + \vec{F}_\mathrm{fr}})/{M_d}
\end{equation}

Equation (1) is written for the vertical coordinate, $z$, which is the height above the horizontal lower electrode. The particle’s mass $M_d$ and radius $R_d$ are related by the mass density $\rho_d$,

\begin{equation}
M_d = \frac{4}{3} \pi \rho_d R_d^3
\end{equation}

The time series for the vertical component of the dust particle’s velocity, $v_{dz}$, is the quantity that we will compare to experimental data.  

We take into account three forces: gravitational force, $M_d\,g$, electric force $F_E$, and gas-dust  friction $F_\mathrm{fr}$. The latter, according to the Epstein theory \cite{Epstein_1924}, is described by

\begin{equation}
\vec{F}_\mathrm{fr}/{M_d = -\delta \, \frac{4}{3} \pi R_d^2 m_n \bar{c} \, N \dot{\vec{v}}_{dz}}/{M_d}
\end{equation}

Combining (2) and (3), the gas friction force is

\begin{equation}
\vec{F}_\mathrm{fr}/M_d = -\delta \, \rho_d^{-1} R_d^{-1} m_n \bar{c} \, N \, \dot{\vec{v}}_{dz}
\end{equation}

In Eq. (2), the gas atoms are described by their mass $m_n$, number density $N$, temperature $T_n$, and mean thermal speed $\bar{c} = \sqrt{8 k_B T_n/\pi m_n}$. We note that $\bar{c}$ is not the same as the thermal velocity \cite{Merlino_2025}. The Epstein drag coefficient $\delta$ can range from 1 to 1.44, according to the Epstein theory \cite{Epstein_1924}, depending on how gas molecules bounce from the surface of the dust particle. The value of $\delta$ can reach its theoretical maximum of 1.44 if the molecules bounce at diffuse angles \cite{Epstein_1924}, which will be a condition of interest later, in our analysis.

To obtain a time series for $v_{dz}$, in the model we integrate Eq.\,(1) numerically. In a test, we found that Euler’s method of integration was adequate for this integration, with a fixed time step of 0.5\,ms. That time step is one-half of the time interval between camera frames in the experiment. This integration yields the model’s time-series for velocity $v_{dz}$.

\begin{figure}
	\centering
	\includegraphics[width=1\columnwidth]{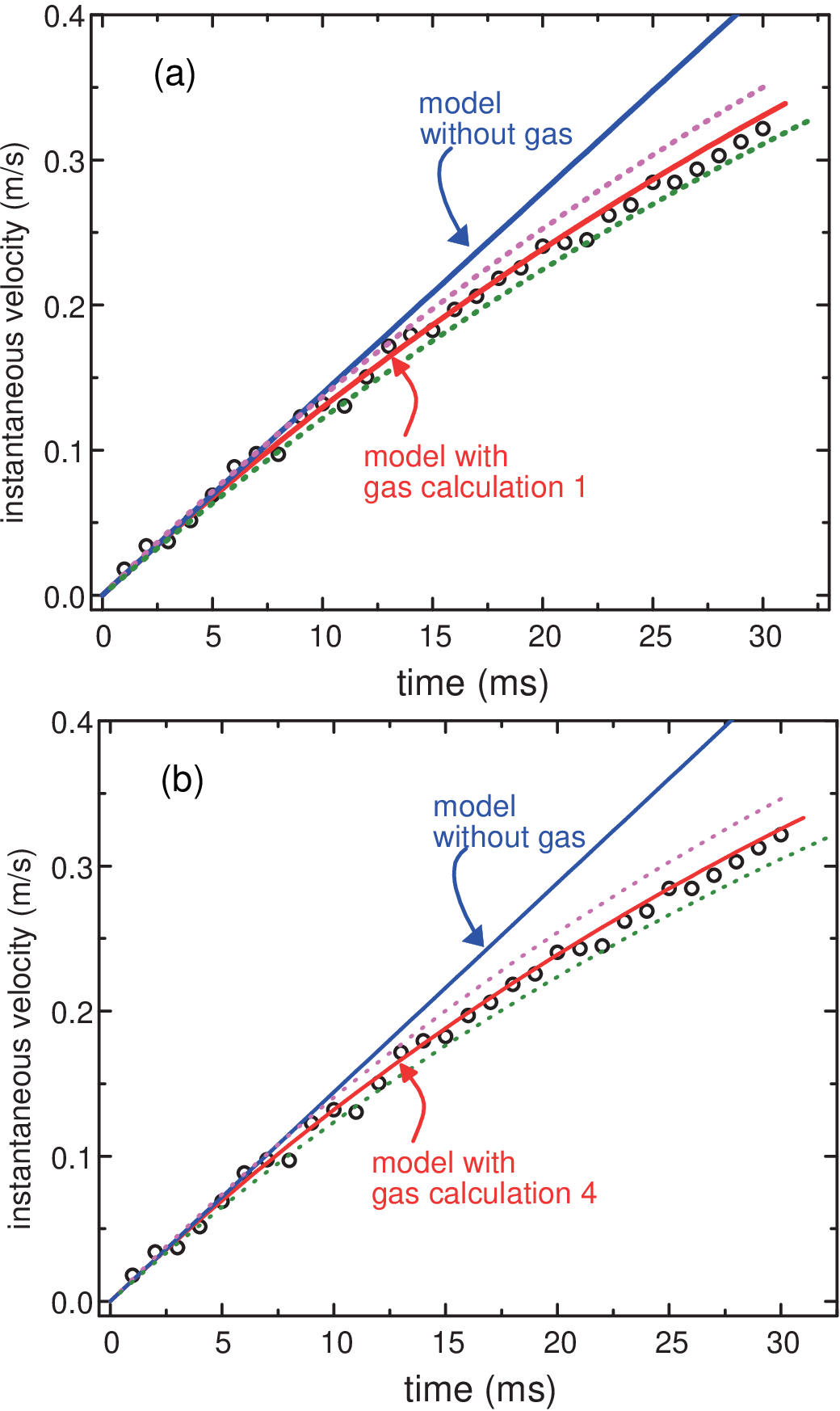}%
	\caption{\label{Fig2} Instantaneous velocity as measured in the experiment (data points) and calculated in the model (curves). Referring to Table I, the input parameters for (a) and (b) were calculations 1 and 4, respectively.  The electric force was our result, obtained as a free parameter to obtain agreement with the experimental data points, yielding  $F_E$ = 2.13\,pN = 0.42\,$M_dg$ for calculation 1, and $F_E$ = 2.4\,pN = 0.72\,$M_dg$ for calculation 4.  This result confirms that even with the higher gas pressure of 90\,mTorr and the large particles used in this experiment, it is possible to manipulate particles in the afterglow by applying a modest DC electric field in the afterglow. As a comparison, to show the effect of gas-dust friction, a straight line is drawn to indicate the uniform acceleration $M_d g$ + $F_E$ that would occur in the absence of gas.  To help gauge random errors, dashed curves bracket the value of $F_E$ by  $\pm\,20\,\%$ in the model. }
\end{figure}

\subsection{Comparison to experiment}
A comparison of the theoretical and experimental time series for $v_{dz}$ allows us to determine the electric force $F_E$. In this comparison, we consider $F_E$ to be a free parameter, which we adjust to obtain the best agreement between experiment and theory. In doing so, we particularly seek to obtain agreement at early times, when the time-series curves were nearly straight lines and gas friction did not yet have much effect. In other words, in adjusting the value of $F_E$, we allow less disagreement between experiment and theory at small times, when the forces acting on the particles were mainly just the downward forces of gravity and $F_E$.  

Random errors in the determination of $F_E$ can arise from variations in our measurements of the individual particles. In particular, the particles do not all have exactly the same height at a given time. To reduce these random errors, we averaged the heights of approximately 30 particles that were imaged within the vertical laser sheet.

Systematic errors in our result for $F_E$ can arise from uncertainties in the three values mentioned earlier, $R_d$, $\rho_d$, and $\delta$. These are inputs to the calculation of the equation of motion that affect the acceleration of particles when the gas friction force is substantial. All three parameters are specific to the sample of particles we used. The manufacturer of our MF microspheres specifies nominal values $R_{d\,\mathrm{mfg}}$ and $\rho_{d\,\mathrm{mfg}}$. However, it has been confirmed by several dusty plasma experimenters \cite{pavlu_2004,carstensen_2011,carstensen_2013, kohlmann2019high} that for MF particles, the actual value of $R_d$ is often smaller than $\rho_{d\,\mathrm{mfg}}$, as the microspheres become smaller with time when exposed to vacuum due to outgassing \cite{pavlu_2004} and when exposed to plasma which causes sputtering or etching \cite{carstensen_2011,carstensen_2013, kohlmann2019high}. The particle mass $M_d$ also has been reported to diminish \cite{pavlu_2004}. While there have been no reported measurements of another parameter, the overall mass density $\rho_d$, we will in our analysis allow the possibility that this quantity can also diminish.  

\subsection{Sensitivity test}

To assess the impact of systematic errors arising from the three uncertain parameters, we will perform sensitivity tests. In these tests, as listed in Table I, we will use a pair of values of each of these ratios: 

\begin{equation*}
{R_d}/{R_{d\,\mathrm{mfg}}} = 1.0~\mathrm{or}~0.9,\\ 
\end{equation*}
\begin{equation*}
\rho_d}/{\rho_{d\,\mathrm{mfg}} = 1.0~\mathrm{or}~0.9,
\end{equation*}
\begin{equation*}
\delta = 1.26~\mathrm{or}~1.44.
\end{equation*}

The second value listed, for each pair, will tend to lead to a greater diminishment of particle acceleration due to gas friction, when calculating using Eq. (1).
  
For the ratio $R_d\,/\,R_{d\,\mathrm{mfg}}$, there are two reasons that we consider a value of 0.9. Firstly, fresh melamine formaldehyde (MF) microspheres typically have diameters smaller than the manufacturer’s specifications, with the discrepancy being typically a few percent, according to electron-microscope measurements by several experimenters as reviewed by \cite{kohlmann2019high}. Secondly, Kohlman $et$\,$al.$ \cite{kohlmann2019high} found that after plasma exposure, the size of the MF microsphere diminishes steadily. In an argon plasma for particular values of RF power and other parameters, they measured the diminishment as $dr/dt$\,=\,-1.25\,nm/min for a particle starting at radius of 4.78\,$\mu$m, so that a $10\,\%$ diminishment of radius can occur in less than one hour of plasma operation, under the conditions that were used by Kohlmann $et$\,$al.$ \cite{kohlmann2019high}. Their chamber was similar to ours, but their operating parameters were presumably different than ours, since we purposefully operated the plasma at nearly the lowest RF voltage that would sustain a plasma. Since we operated our plasma for several hours, it is reasonable to estimate that the radius change may have been of order $10\,\%$ in our experiment.

For the ratio $\rho_d/\rho_{d\mathrm{mfg}}$, we chose 0.9 as the lower of our two values, based on two reports.  Carstensen $et$\,$al.$ \cite{carstensen_2011} reported that $11\%$ of particle mass was water, which outgassed gradually. Pavlu $et$\,$al.$ \cite{pavlu_2004} reported similar results earlier. Our particles were exposed to vacuum for hundreds of days before the experiment, so that it is reasonable to assume that the mass density may have reduced to as little as $\rho_d / \rho_{d\mathrm{mfg}}  \approx 0.9$. (For the purposes of senstivity tests, we will assume that  the evaporation was accounted for only by a change in $\rho_d$, without a change in $r$, although neither Carstensen $et$\,$al.$ \cite{carstensen_2011,carstensen_2013} nor Pavlu $et$\,$al.$ \cite{pavlu_2004} reported that detail.)

For the Epstein drag coefficient $\delta$, the theoretical range of values is $1 \leq \delta \leq 1.44$ for spherical particles. We will consider two values: 1.26 as reported by \cite{Liu_2003} for MF microspheres, and 1.44 corresponding to diffuse reflection of gas molecules in the Epstein theory \cite{Epstein_1924}. Diffuse reflection is a reasonable possibility, especially after the microsphere’s surface has become roughened due to plasma exposure, as has been reported by experimenters \cite{karasev2020mechanism}.

\section{Results}
The experimental velocity time series is shown as symbols in both panels of Fig. 2. For these data, the transistor switch was used to bias the lower electrode positively, at +150\,V, starting at $t$\,= 250\,$\mu$s. The particles fell, requiring about 31\,ms to impact the electrode.

To analyze the experimental time series for particle velocity, we numerically integrated the equation of motion, Eq.\,(1), to yield a velocity time series for the model. We repeated this integration for four different sets of assumptions for the values of the $R_d$, $\rho_d$, and $\delta$. These four sets of values are identified as calculations 1 through 4 in Table I, where we normalize $R_d$ and $\rho_d$ by the nominal values specified by the manufacturer, $R_{d\,\mathrm{mfg}}$ and $\rho_{d\,\mathrm{mfg}}$.

In Fig.\,2 we present as solid curves the results for the model time series for $v_{dz}$.  In Fig.\,2(a), the solid curve is marked as “model with gas calculation 1,” and for this calculation we assumed  $R_d$ / $R_{d\,\mathrm{mfg}}$,  $\rho_d$\,/\,$\rho_{d\,\mathrm{mfg}}$, and $\delta$ were 1, 1, and 1.44, respectively.  For this solid curve, the result we obtained for the free parameter was $F_E$ = 2.13\,pN, for agreement with the experimental data points. This downward force, expressed in terms of the gravitational force $m_dg$, is $F_E$\,=\,0.42\,$m_dg$. In Fig.\,2(a), dashed curves bracket this value of $F_E$ by  $\pm\,20\,\%$, allowing us to judge from the scatter of experimental data points that their random errors were less than $\pm\,20\,\%$.

In our sensitivity test, to assess systematic errors arising from the three input values ($R_d$, $\rho_d$ and $\delta$), we repeated our analysis using the four sets of input values listed in Table I. The fourth set, in what we call calculation 4, yielded the best agreement with the experiment, as seen in Fig.\,2(b). This particular set of values maximized the gas friction force in comparison to gravity, i.e., its value of $F_\mathrm{fr}$\,/\,$M_d g$\,=\,0.72\,$m_dg$ was larger than the other three sets. This comparison to the other sets indicates that it is very likely that the particle mass is less than one would expect from using the manufacturer’s nominal specifications in Eq.\,(2). 

\begingroup            
\setlength{\tabcolsep}{4pt}   
\renewcommand{\arraystretch}{0.95}  

\begin{table}[h!]
\centering
\footnotesize  
\caption{Results for the electric force $F_E$, showing that it is possible to control the lifting of particles. The four rows, labeled calculations 1-4, are for a sensitivity test to assess systematic errors arising from uncertainties in the particle parameters $R_d$ , $\rho_d$ and $\delta$ , which are inputs in the equation of motion. The particle mass $M_d$, which we calculated using Eq.\,(2) is another input to the equation of motion. Comparing results for $F_E$ for these four calculations indicates that systematic uncertainties are small enough to allow concluding that it is possible to apply a significant electric force, even at the higher gas pressure and large particles used in this experiment.}
\begin{tabular}{|c|c|c|c|c|c|c|}
\hline
\textbf{Calc.} &
\multicolumn{3}{c|}{\makecell{\textbf{Inputs for}\\\textbf{Equation of Motion}\\(dimensionless)}} &
\textbf{Mass} &
\multicolumn{2}{c|}{\textbf{Results}} \\
\cline{2-4}\cline{6-7}
 & $R_d / R_{d\mathrm{mfg}}$ & $\rho_d / \rho_{d\mathrm{mfg}}$ & $\delta$ &
 $M_d$ (pg) & $F_E$ (pN) & $F_E / M_d g$ \\
\hline
1 & 1.0 & 1.0 & 1.44 & 520 & 2.13 & 0.42 \\
2 & 1.0 & 1.0 & 1.26 & 520 & 1.99 & 0.39 \\
3 & 1.0 & 0.9 & 1.44 & 468 & 1.71 & 0.37 \\
4 & 0.9 & 0.9 & 1.44 & 342 & 2.40 & 0.72 \\
\hline
\end{tabular}
\end{table}

Examining Table I, we can conclude that there is no doubt that the downward electric force $F_E$ was substantial in this experiment. It was a substantial fraction of the gravitational force, even for the large particle size that we used.
  
\section{Discussion}
\subsection{Summary of experiment}
We performed an experiment with afterglow charging of dust particles, for the purpose of extending the parameter range that the charge can be controlled. We operated with a higher gas pressure of 90\,mTorr, as compared to 8 mTorr in our previous experiments \cite{Neeraj_1,Neeraj_2,Neeraj_3,Neeraj_4,Neeraj_5,Neeraj_6}.  Particles fell when we turned off the RF power that sustained that plasma, and after a delay of 250\,$\mu$s we applied a DC potential of +150\,V to the lower electrode in order to reverse the electric field direction, at a time when electrons and ions were still abundant in the afterglow.

To obtain our main result, demonstrating that the electric force $F_E$ can be controlled in the afterglow, even at this higher gas pressure, we measured the velocity time series of the particles as they fell, and compared to a theoretical calculation. That calculation involved integrating the particle’s equation of motion, taking into account three forces: gravity, gas drag, and $F_E$, where the latter quantity was adjusted in the calculation to obtain agreement with experiment. We determined that random errors were not significant. Systematic errors can also occur, especially from diminishment of the particle size and density, due to exposure to plasma and vacuum. We carried out a sensitivity test, in our analysis, and determined that even when taking into account these systematic errors, we can confirm that $F_E$ was a large fraction of the gravitational force. This result confirms that it is possible to manipulate particles using purposeful application of electric fields in the afterglow, even at the higher gas pressure of the present experiment.

Aside from that major result, we also gained information about the particular particles in our experiment. For these MF microspheres, we determined that their diameter and mass were likely smaller than the manufacturer’s specifications. We also determined that the Epstein drag coefficient $\delta$ has a value that is as large as 1.44. The latter result makes sense because it is known that plasma exposure causes an MF particle’s surface to become roughened, which we expect would tend to cause gas molecules to bounce diffusely from the particle’s surface. Diffuse reflection corresponds to $\delta$\,=\,1.44 in the Epstein theory \cite{Epstein_1924}.

\subsection{Application to semiconductor manufacturing}
Manipulation of dust particles in a plasma afterglow is of interest in semiconductor manufacturing. After turning off the power that sustains the plasma, particles would normally fall downward, where they can contaminate critical surfaces like wafers and reticles. By purposefully applying upward forces during the afterglow, the falling of these particles can be slowed, or reversed entirely, providing more time for gas flow to purge particles and thereby reducing contamination \cite{Neeraj_5}. 

In this paper, we extended the parameter range for which we have confirmed that manipulation of particles should be possible. We performed an experiment that demonstrates that it is possible to apply substantial electric forces to a dust particle in an afterglow plasma, even at gas pressures that are elevated well above the 8\,mTorr level of our previous experiments \cite{Neeraj_1,Neeraj_2, Neeraj_3, Neeraj_4, Neeraj_5, Neeraj_6}. Using argon at 90\,mTorr, we attained an electric force that was nearly as strong as gravity, even when using large 8.69\,$\mu$m diameter particles. We did this by applying a modest positive potential of 150 V to the lower electrode 250\,$\mu$s after turning off the RF power, so that the dust particles near the lower electrode were immersed only in electrons during the remaining time of the afterglow. This application of a DC potential to manipulate particles for the purpose of avoiding contamination was described in our patent applications \cite{Neeraj_7,Neeraj_8}. 

Higher gas pressure has two consequences in the afterglow. First, it slows the drifting motion of electrons or ions in the afterglow. Second, it slows the fall of particles. For the purpose of manipulating particles, the first effect is of interest: electrons and ions undergo mobility-limited motion in the presence of a DC electric field that is applied in the afterglow, so that a higher gas pressure reduces the kinetic energy of electrons and ions in the afterglow. We showed that in an afterglow, when a dust particle is immersed only in electrons, those electrons can still charge the dust particle sufficiently, even at a higher gas pressure of 90\,mTorr, thereby allowing electric manipulation of the dust particles in the afterglow.

Extrapolating our results to smaller particle sizes is possible because of the known scaling of forces.  This is of particular interest for the semiconductor industry, where contaminating particles are generally smaller than the ones that we used. We expect that it will be even easier to manipulate smaller particles, because of the way that forces scale with particle size. The scaling with particle radius  $R_d$ is $\propto$ $R_d ^3$ for gravity, and $\propto$ $R_d ^1$ for electric forces. (The gas drag force scales as $\propto R_d ^2$ for gas drag, but that is of less interest here.) Thus, the ratio of forces $F_E$\,/\,$m_dg$ will tend to scale $\propto R_d ^{-2}$. 

Because of that scaling, a ten-fold diminishment of particle size would lead to a thousand-fold increase in the ratio $F_E$\,/\,$M_d\,$g. Thus, for smaller particles, electric forces more easily dominate gravity. Even for the large particles in our experiment, we were able to obtain a substantial electric force of magnitude $F_E$\,/\,$M_d\,g$ in the range 0.42 to 0.72 merely by applying a modest potential of 150\,V to the lower electrode during the afterglow. It would be even easier to exceed the force of gravity when the particles are smaller. 

If the gas pressure were raised higher than the 90\,mTorr level that we used, the dust particles would charge to a lesser potential, due to a slower motion of electrons (or ions) in the afterglow. This effect, which would reduce the electric force, could be overcome by applying a greater electric potential to the lower electrode.  On the other hand, we note that submicron particles would experience an even larger effect than we observed, so that for such particles, even a gas pressure of 1\,Torr or higher would pose no obstacle to the practical manipulation of their motion.

\section*{Data Availability}
The data that supports the findings of this study are available within the article.

\ack
Work at Iowa was supported by the U.S. Department of Energy grant DE-SC0025444 and National Science Foundation grant (No. PHY-2510501).

\section*{References}
\bibliographystyle{iopart-num}
\providecommand{\noopsort}[1]{}\providecommand{\singleletter}[1]{#1}%
\providecommand{\newblock}{}

\end{document}